\documentclass[showpacs,preprint,aps]{revtex4}
\usepackage{graphicx}
\usepackage{dcolumn}
\usepackage{bm}
\usepackage{amssymb}
\usepackage{amsmath}
\begin{document}
\setcounter{page}{1}
\title 
{The fate of a gravitational wave in de Sitter spacetime}
\author
{M. Nowakowski and I. Arraut}
\affiliation{ 
Departamento de Fisica, Universidad de los Andes, 
Cra.1E No.18A-10, Bogota, Colombia}
\begin{abstract}
If we want to explain the recently discovered accelerated stage of the
universe, one of the
option we have is to modify the Einstein tensor. 
The simplest such modification,
in agreement with all observations, is the positive cosmological constant
$\Lambda$.
Such a modification will also have its impact on local observables and on
the propagation of weak gravitational waves. We show here that
the inclusion of a cosmological constant impedes the detection of a
gravitational wave if the latter is produced at a distance larger than
${\cal L}_{\rm crit}=(6\sqrt{2}\pi f \hat{h}/\sqrt{5})r_{\Lambda}^2$ where
$r_{\Lambda}=1/\sqrt{\Lambda}$ and $f$ and $\hat{h}$ are the 
frequency and the strain of the wave, respectively. ${\cal L}_{\rm crit}$ is
of astrophysical order of magnitude. We interpret the result
in the sense that the gravitational wave interpretation is only possible if 
the characteristic wave properties are smaller than the
non-oscillatory solution due to $\Lambda$.
\end{abstract}
\maketitle 
\section{Introduction}
In testing Einstein's theory of gravity, its modifications and ramifications, 
two important sub-areas of research remain to be explored and explained
in more detail. The first one has to do with cosmology and goes back to 
the discovery of dark energy ten years ago 
which drives the acceleration of the universe
\cite{darkenergy, darkenergyexp}. The second one is the possibility
to detect gravitational waves \cite{gravwave} directly \cite{hulsetaylor} by
already operating \cite{ligo} or forthcoming 
\cite{lisa, etc} gravitational wave detectors.
In order to explain the accelerated universe, we can either modify
the Einstein's tensor $G_{\mu\nu}$ 
or try to suitably alter the cosmological energy-momentum tensor.
The first category encompasses modified gravity theories
and theories with the inclusion of a positive cosmological constant $\Lambda$ \cite{lambda}.
This simplest modification is in agreement with all observations and, notably,
its equation of state $\rho=-p$ is observationally also favored \cite{eoslambda}.
Once we change the Einstein's tensor to explain cosmological facts, 
we are
also forced to accept the fact that the very same parameters will also
affect local physics, at least in principle. Hence, for instance, the
Schwarzschild metric becomes Schwarzschild-de Sitter metric where
$\Lambda$-effects are also sizable on local scales. Indeed, the 
theory now contains two lengths scales, the small Schwarzschild radius
$r_s$ and the large 
cosmological scale $r_{\Lambda}=1/\sqrt{\Lambda} \sim H^{-1}_0$ where
$H^{-1}_0$ is the Hubble radius. However, a combination like
$(r_sr_{\Lambda}^2)^{1/3}$ is of astrophysical order of magnitude and has
the meaning to be the distance beyond which no bound orbits are 
possible \cite{deSitter}. Other local effects of the cosmological constant 
can be found in \cite{we2rest}. Similarly, the linearized
version of the now modified, Einstein's equations will include the
cosmological constant. These expressions are not any more interpretable
as the Fierz-Pauli equations \cite{FierzPauli} for a spin-2 object.
Nonetheless we can still
understand them as a mathematical approximation for weak fields.
Moreover, part of these linearized equations will contain
the Fierz-Pauli term and therefore
the question about gravitational waves in the new theory can be also 
addressed in a meaningful way. What remains to see is how exactly the modification of 
Einstein's tensor influences the propagation of the gravitational waves.
To this end we solve the linearized equations with $\Lambda$ and use them 
in the energy-momentum pseudo-tensor of gravity to study the
effect of the cosmological constant. The result, which 
can be formulated in form of a critical distance, is proportional
to $r_{\Lambda}^2$ and depends on
the frequency and amplitude of the wave and the distance
of the source from the detector. Although $r_{\Lambda}$ is of cosmological
order  of magnitude, the small amplitude of the wave arriving
on earth renders it possible that the negative contribution of $\Lambda$ 
to the power $P$ is as large as the standard oscillatory one. 
As a result, though the promises to detect gravitational waves 
from most of the systems can be fulfilled, there are some whose
gravitational waves detection is impeded by $\Lambda$.
This is a curious effect of the accelerated universe which gives us also the
chance to probe the theory of dark energy through gravitational wave 
detection, provided we know the exact distance of the source.
The interpretation of our result involves 
a distinction between two tests for gravitational waves: (i) an internal
comparison between the two parts of the solution 
(wave and the non-oscillatory part) and (ii)
a comparison between the wave 
solution and the cosmological background.
\section{Linearized Einstein's equation with $\Lambda$}
We start with the linearized Einstein's equations with $\Lambda$
for weak field $h_{\mu\nu}$ i.e. the metric is $g_{\mu\nu}=\eta_{\mu\nu}+h_{\mu\nu}$
\cite{footnote}
where $\eta_{\mu\nu}$ is the Minkowski metric (our conventions are like in \cite{weinberg}):
\begin{equation} \label{Einstein}
R_{\mu \nu}^{(1)}=-8\pi GS_{\mu\nu}-\Lambda \eta_{\mu \nu}
\end{equation}
where we have used the trace-reversed part of the energy-momentum tensor
\begin{equation} \label{S}
 S_{\mu \nu}\equiv T_{\mu \nu}-\frac{1}{2}\eta_{\mu \nu}T.
\end{equation}
The linearized expression of the Ricci tensor is easily obtained to be
\begin{equation} \label{Ricci1}
R^{(1)}_{\mu \nu}\equiv \frac{1}{2}(\square h_{\mu \nu}
-\partial^\lambda \partial_\mu h_{\lambda \nu}-\partial^\lambda \partial_\nu h_{\lambda \mu}
+\partial_\mu \partial_\nu h)
\end{equation}
which gives us the linearized equations 
\begin{equation} \label{lineareq}
\square h_{\mu \nu}-\partial^\lambda \partial_\mu h_{\lambda \nu}
-\partial^\lambda \partial_\nu h_{\lambda \mu}+\partial_\mu \partial_\nu h
=-16\pi GS_{\mu \nu}-2\Lambda \eta_{\mu \nu}.
\end{equation}
This equation is clearly covariant under the local gauge transformation $h_{\mu \nu}\to h_{\mu \nu}
+\partial_\mu \epsilon_\nu+\partial_\nu \epsilon_\mu$ as imposed by the general diffeomorphic covariance
of the Einstein's equations with $\Lambda$.
Any attempt to make the cosmological constant more dynamical by
replacing $\Lambda \eta_{\mu\nu} \to \Lambda g_{\mu\nu}$ would violate this gauge covariance (in the Appendix
we discuss this issue in more detail and confirm the validity of (\ref{lineareq})).
This gauge freedom allows us
to fix the gauge which we choose to be the de Donder condition: 
$\partial^\mu h_{\mu \nu}=\frac{1}{2}\partial_\nu h$. 
The equation to be solved becomes a wave equation with two kinds of inhomogeneities; 
one the standard source
$S_{\mu\nu}(x)$, the other one a constant term proportional to the cosmological constant:
\begin{equation} \label{waveeq}
\square h_{\mu \nu} =-16GS_{\mu \nu}-2\Lambda \eta_{\mu \nu}.
\end{equation}
Since the equation is linear we can split its solution $h_{\mu\nu}$ in two parts
\begin{equation} \label{solutiongeneral}
h_{\mu\nu}=\gamma_{\mu\nu}+\xi_{\mu\nu}.
\end{equation}
where $\gamma_{\mu\nu}=e_{\mu\nu}({\bold r}, \omega)e^{ik_{\alpha}x^{\alpha}} 
+ {\rm c.c.}$ is the standard retarded solution (written here 
for a monochromatic source at a distance far away from the source \cite{weinberg}) and $\xi_{\mu\nu}$
solves $\square \xi_{\mu\nu}=-2\Lambda \eta_{\mu\nu}$. The latter, 
should satisfy the de Donder gauge and, in addition, we demand that
up to a diffeomorphism its asymptotic form is of the de Sitter metric.
Both the conditions fix the constants $a$ and $b$ 
and the solution of the homogeneous wave equation
$\xi_{\mu\nu}^{(2)}$ (this is necessary to
satisfy the de Donder condition) in the general ansatz
$\xi_{\mu\nu}^{(1)}+\xi_{\mu\nu}^{(2)}$
where $\xi_{\mu\nu}^{(1)}=(ar^2 +bt^2)\eta_{\mu\nu}$. 
In other words, $\xi^{(1)}_{\mu \nu}$ is the initial ansatz, 
supplemented by $\xi^{(2)}_{\mu \nu}$ which guarantees that
the metric is asymptotically de Sitter, 
and the de Donder condition is satisfied. 
The full solution which is in agreement with \cite{dvali} (
to compare with \cite{dvali} one has to take the
graviton mass $m$ to $0$ in \cite{dvali}) reads,
\begin{equation}\label{solutionLambda}
\xi_{0 0}=-\Lambda t^2, \,\,
\xi_{0 i}=\frac{2}{3}\Lambda tx_i,\,\,
\xi_{i j}=\Lambda t^2\delta_{i j}+\frac{1}{3}\Lambda\epsilon_{i j},
\end{equation}
where $\epsilon_{i j}=x_i x_j$ for $i\neq j$ and $0$ otherwise.
These solutions will be used in the energy momentum pseudo-tensor
$\hat{t}_{\mu\nu}$
for gravitational waves. 

\section{The energy momentum tensor}
In the absence of the 
cosmological constant the latter is defined as $(G_{\mu\nu}-
G_{\mu\nu}^{(1)})/8\pi G$ \cite{LandauLifschitz} where , again, the index $(1)$ indicates
that we expand the tensor in the order ${\cal O}({\bf h})$. Taking
into account that $G_{\mu\nu}$ is now modified, the very same
procedure can be adopted for theories with $\Lambda$
leading to
\begin{equation} \label{tmunu} 
\hat t_{\mu \nu}=t_{\mu \nu}-\frac{1}{8\pi G}\Lambda h_{\mu \nu} 
\end{equation}
where $t_{\mu \nu}$ is the part defined by

\begin{equation}   \label{eq:self-energy-momentum}
t_{\mu \nu}=\frac{1}{8\pi G}\left(-\frac{1}{2}h_{\mu \nu}R^{(1)}
+\frac{1}{2}\eta_{\mu \nu}h^{\sigma\rho}R^{(1)}_{\sigma\rho}
+R^{(2)}_{\mu \nu}-\frac{1}{2}\eta_{\mu \nu}\eta^{\sigma\rho}R^{(2)}_{\sigma\rho}\right)
+{\cal O}({\bf h}^3).
\end{equation}

Note that $t_{\mu\nu}$ is of the order ${\bf h}^2$. In agreement with the linearized equation
of motion which is of the first order in ${\bf h}$ and first order in $\Lambda$, the effects
of the cosmological constant in gravitational waves can be only of the order $\Lambda^2$ or
$\Lambda {\bf h}$. 
It remains to calculate 

\begin{equation}   \label{toi}
\hat{t}_{0 i}=\frac{1}{8\pi G}\left(-\frac{1}{2}h_{0 i}\eta^{\lambda \rho}R^{(1)}_{\lambda \rho}
+R^{(2)}_{0 i}-\Lambda h_{0i}\right)+{\cal O}({\bf h}^3)
\end{equation}

which in the averaged form $< \hat{t}_{0i}>$
enters the expression for the power of the gravitational waves. Making use of the equation of motion
$R^{(1)}_{\mu \nu}=-\Lambda\eta_{\mu \nu}$,
we obtain three contributions of $\Lambda$ to the gravitational Poynting vector $\hat{t}_{0i}$, namely
\begin{equation}   \label{toi2}
\hat{t}_{0 i}=\frac{1}{8 \pi G}\left(\frac{4}{3}+\frac{4}{9}-\frac{2}{3}\right)\Lambda^2tx_i+...      
= \frac{1}{8\pi G}\left(\frac{10}{9}\Lambda^2 t x_i\right)+...
\end{equation}
indicating the different contributions in the same order as in equation (\ref{toi}).
The dots stand for oscillatory contributions proportional 
to \mbox{\boldmath $\gamma \xi$} (which average to zero)
and the standard terms proportional to 
$\mbox{\boldmath $\gamma$}^2$ surviving the averaging process.
The explicit calculation of the contribution $R_{0i}^{(2)}$ is lengthy albeit straightforward. 
Assuming the direction of the wave to be
$z$, the important quantity for us is $<\hat{t}^{03}>=<\hat{t}^{03}>_{\rm wave}
+<\hat{t}^{03}>_{\Lambda}$ where the subscript `wave' refers to the standard
contribution without the cosmological constant. Taking into account that
the wave-front moves with the velocity of light (which entitles us to 
identify time with the distance $L$) one calculates
\begin{equation} \label{average}
<\hat{t}^{03}>_{\rm wave}=<t^{03}>_{\rm wave}=\frac{\omega^2 \hat{h}^2}{8\pi G},
\,\,\, <\hat{t}^{03}>_{\Lambda}=-\frac{1}{8\pi G}\frac{5}{18}\frac{1}{r^4_{\Lambda}}L^2,
\end{equation}
where $\hat{h}$ is either $\vert e_{11} \vert$ or $\vert e_{12} \vert$.
Note that due to $\Lambda$, the power
\begin{equation} \label{power}
\frac{dP}{d\Omega}=r^2\frac{x_i}{r}<\hat t^{0 i}>
\end{equation}
receives a negative contribution. The power is only well defined i.e. positive definite 
below a certain critical distance ${\cal L}_{\rm crit}$ 
where the oscillatory character of the solution dominates.
To calculate this critical distance it suffices to compare the magnitudes of the
two contributions to $<\hat{t}^{03}>$. The result is
\begin{equation} \label{critical}
{\cal L}_{\rm crit}=\frac{6\sqrt{2}\pi f\hat{h}}{\sqrt{5}} r^2_{\Lambda}.
\end{equation}
Had we not modified the energy-momentum gravitational pseudo-tensor 
$t_{\mu\nu}$ to become
$\hat{t}_{\mu\nu}$ in equation (\ref{tmunu}) the contribution of $\Lambda$ would be bigger
and the critical distance smaller by  a factor $0.8$ which would increase its relevance.
In interpreting the  above result we emphasize that there is little doubt that
a modification of the Einstein's tensor will change the linearized version of the
Einstein's equation (in our case with $\Lambda$ this is equation (\ref{lineareq})). One could also
paraphrase this in saying that the Newtonian Limit will change \cite{barrowme}. As a consequence,
the solution will now contain an oscillatory and a new contribution originating in
the modifications of $G_{\mu\nu}$ (proportional to $\Lambda$ in our case).
The interpretation of gravitational waves as ripples on spacetimes can be only maintained
if the oscillatory solution is more sizable than the non-oscillatory one proportional to $\Lambda$.
The result in (\ref{critical}) reflects exactly this fact. 
One can also view this result from a more formal perspective. Even though the cosmological constant
is not part of the energy-momentum tensor, one can nevertheless, formally, absorb it there as
evident in (\ref{lineareq}). It is then obvious that $\Lambda$ will be always a source for the metric, 
gravitational waves to which it contributes, not excluded.

Notice that what we are really comparing is the averaged solution 
proportional $\Lambda$ with the averaged wave component of the solution. We
then say that the wave character of the solution is lost when both are 
comparable.

\begin{table}[h]
\caption{\label{tab1} Sources of gravitational waves for LIGO from references \cite{ligoobs}. 
AIC means accretion induced collapse.
For $P$ we have used
geometrized units $G=c=1$.}
\begin{tabular}{|l|l|l|l|l|l|l|}
\hline
System 
& f [Hz] & $\hat{h}$
&Distance & ${\cal L}_{\rm crit}$& $dP/d\Omega$& 
$dP/d\Omega$\\
 & &  &[pc] & [pc] & $\Lambda=0$& $\Lambda \neq 0$ \\ 
\hline
\hline
NS/NS binary &$100$  &$1\times 10^{-23}$  &$10^{9}$ &$12.9\times 10^{6}$  &$-$ &$-$   
\\ \hline
BH/BH binary &$100$  &$1\times 10^{-22}$  &$2\times 10^8$ &$12.8\times 10^7$  &$-$ 
& $-$
\\ \hline
Collapse and explosion & $20$ &$4.1\times 10^{-23}$  
&$10^{7}$ &$10.6\times 10^{6}$   &$1.11\times 10^{-12}$ & $1.18\times 10^{-13}$ \\ 
of Supernova & & & & & &\\
\hline 
NS formed  &$450$  &$8\times 10^{-23}$  &$10^8$ &$46\times 10^7$  &$2.15 \times 10^{-7}$ 
& $2.05 \times 10^{-7}$\\
from AIC & & & & & &  \\ \hline 
NS/NS binary &$1000$  &$1\times 10^{-20}$  &$2.3\times 10^7$ &$12.8\times 10^{10}$  &$8.79 \times 10^{-4}$ 
& $8.79 \times 10^{-4}$
\\ \hline 
Stellar collapse &$100$  &$1\times 10^{-22}$  &$15\times 10^6$ &$12.9\times 10^7$  &$3.74 \times 10^{-10}$ 
& $3.69 \times 10^{-10}$\\
Centrifugal hang up  & & & & & &  \\ \hline 
\end{tabular}
\end{table}
\noindent
\section{Phenomenological results}
In exploiting (\ref{critical}) phenomenologically, we first point out that the gravitational waves
arriving on earth are indeed weak which is exactly the reason making their detection difficult.
They can also be considered weak over the largest part of the distance they travel to earth.
Therefore, even if (\ref{critical}) is an approximation, it is a rather good one. 
Secondly, the very same fact that the waves arriving are weak  makes ${\cal L}_{\rm crit}$
of astrophysical order of magnitude in spite of the large value of $r_{\Lambda}$.
To see that, let us take some typical values: $f=0.38 \times 10^{-3} {\rm Hz}$ and $\hat{h}=40 \times
10^{-23}$. We obtain ${\cal L}_{\rm crit}= 1957 {\rm pc}$. The values taken are for the white dwarf
binary system WD 0957-666 whose distance from earth is expected to be roughly $100{\rm pc}$.
In this case even though the critical distance is of astrophysical order of magnitude,
the gravitational waves from the white dwarf devil's system will be seen as its
distance from earth is smaller that the critical one.  
The detector sensitive to the values of frequency and amplitude (strain) would be in this case 
the forthcoming space-located LISA detector \cite{lisa}. 
Another example is the collapse of rotating stare cores \cite{Dimmelmeier} suited e.g. for the
planned Euro detector \cite{Euro}. With the characteristic amplitude $h_c\simeq 10^{-24}$, the
frequency $f\simeq 900{\rm Hz}$ and the relation $h_c=\sqrt{\pi}f\tau\hat{h}/2$ \cite{Abbot}
with $\tau\simeq 10^{-3}{\rm s}$ the duration time, we obtain ${\cal L}_{\rm crit}=1.16{\rm Mpc}$. 
The maximally allowed distance from earth is supposed to be  $d=15{\rm Mpc}$, 
which implies that the range for the gravitational waves to be 
detected is much smaller than $d$  
if $\Lambda$ enters the Einstein's equations with the value needed to explain
the accelerated universe. 
In tables I and II we have listed three kinds of examples for the LIGO and
LISA detectors, respectively. 
In examples where the wave is not monocromatic, we pick up one frequency and the corresponding
amplitude.
The first two entries serve the purpose to demonstrate that
indeed according to (\ref{critical}) the detection of some gravitational waves will be
impeded by $\Lambda$. The next two examples show that the two contributions to the gravitational
Poynting vector can be of the same order of magnitude reducing thereby the power of 
the gravitational wave.
Finally, and this is the majority of cases, the last two examples show that the effect 
of $\Lambda$ can 
be also negligible. This allows us to conclude that constructing a more exhaustive map of 
all sources whose
gravitational waves will not be seen on earth, provided the cosmological constant 
is the right explanation of
the accelerated universe, is a worthwhile undertaking.   
Maybe in the near future we will enjoy to see the
connection between dark energy and gravitational waves
which is not only important for the latter, but converts the gravitational wave detectors partly also
in experimental devices to check dark energy.
A better knowledge of the distance of the source is here a crucial ingredient and
would require an improvement. Indeed, the $100{\rm pc}$ which appears so often in table
II seems to be only an order of magnitude estimate. If its estimate goes up by  a factor $2-5$,
several sources might fall into the category whose gravitational waves will not be see due to $\Lambda$.
Thus the good knowledge of characteristics of the source are of utmost importance for the critical
distance. 
\section{Interpretation}
In this section we argue that due to the appearance of $\Lambda$ in the Einstein
tensor (and not in the energy-momentum tensor) {\it two} tests of gravitational waves are required.
The first one (the cosmological test) is more standard and is due to the interpretation of the gravitational waves
as ripples on spacetime. Here $\Lambda$ appears  in the solution of the cosmological background.
The other test based on (\ref{critical})  is {\it between two solutions}: the oscillatory part versus the
non-oscillatory proportional $\Lambda$. It appears at the first glance that this is the same, 
especially as at present epoch our universe is dominated by $\Lambda$. To see that these two
tests are different, imagine a universe with a non-zero cosmological constant (say, of the same value
as in our universe) where, however, the cosmology is dominated by the background density (i.e. we can
neglect $\Lambda$ is the cosmological equations). The crucial point is now that the second test
relying on (\ref{critical}) would be still required and its outcome would be just the same as
presented in the tables. The effect of Dark Energy models which modify the cosmological
fluid equation (i.e. the ingredients of the model enter the energy-momentum tensor and not the
Einstein tensor) could be probed only in a cosmological test which is not what we have done here.
This is also true for effects which rely on general equation of state distinguishing the Dark Energy
models \cite{boyle}. Such a distinction does not make a difference if the model modifies the Einstein tensor
or the energy-momentum one.

\subsection{$\Lambda$ in cosmology and local physics}
We seem to be biased by the the name ``cosmological constant''
which has instilled in some the impression
that $\Lambda$ is good for cosmology and nothing else. 
As mentioned in the Introduction, this is not correct, but it makes sense
to look at it from a different perspective. 
Consider, for instance, a gravity theory defined by the more
general action
\begin{equation} \label{f}
S=\kappa \int d^4x \sqrt{-g}f(R)
\end{equation} 
where $f(R)$ is more complicated than the standard $f(R)=R$.
Any new constant which appears in $f(R)$ will also appear
in the calculation of any effect of local physics. It would be hard to
argue that no other constant apart from the Newtonian one ($G_N$)
can enter local physics if the gravitational theory is described by 
(\ref{f}). Equally, it would be hard to argue
that no constant other than $G_N$ can affect the solution
of the linearized versions or that the effects due to the new constant are
coordinate effects. 
In view of (\ref{f}) we would be forced to interpret the
gravitational waves anew since part of the solution would 
involve the new constants entering $f(R)$. 
The situation 
with $\Lambda$ is just a special case of what we outlined above.
What is really required in  
cases where the standard $G_{\mu\nu}$ is modified, 
is to pay attention to the interpretation of
gravitational waves. 

First let us note that no local effect of $\Lambda$ is 
per se a coordinate effect.
One cannot get rid of the cosmological constant $\Lambda$
in deriving effects on local physics, in general, and in
the linearized version of Einstein's equation, in particular.
$\Lambda$ is 
an integral part of the modified Einstein's tensor
$G_{\mu \nu}$ and not of the energy-momentum tensor. 
This implies that $\Lambda$ will appear locally, in principle,  everywhere 
where gravitational effects are considered.  More specifically, we have
two equations. The first, 
\begin{equation} \label{e1}
G_{\mu\nu}=R_{\mu\nu}-\frac{1}{2}g_{\mu\nu}R+\Lambda g_{\mu\nu}
=\kappa T_{\mu\nu}^{\rm Universe} 
\end{equation}
for the universe defines the cosmological background and the 
Hubble flow and the second one where $\Lambda$
affects the expansion of the universe, and
\begin{equation} \label{e2}
G_{\mu\nu}=R_{\mu\nu}-\frac{1}{2}g_{\mu\nu}R+\Lambda g_{\mu\nu}
=\kappa T_{\mu\nu}^{\rm Local} 
\end{equation}
for the local physics including the gravitational waves. In spite of the fact that
$\Lambda$ appears already in the cosmological part (\ref{e1}),
it makes its entry once again in calculating local effects.
In a concrete example, a star is part of the Hubble flow (expansion)
where $\Lambda$ already contributes, nevertheless the Schwarzschild-de Sitter
metric (see e.g. \cite{Rindler}) will again contain terms with
$\Lambda$ which is a local effect of this constant.
Another way to see it, is the Newtonian Limit.
$\Lambda$ survives the Newtonian Limit (\cite{Rindler, barrowme}) 
\begin{equation} \label{NL}
\Phi=-\frac{r_s}{r}-\frac{1}{6}\frac{r^2}{r_{\Lambda}^2},\,\,
r_s=G_NM,\,\, r_{\Lambda}=1/\sqrt{\Lambda}
\end{equation}
because locally its effects are not coordinate effects.
Note that the background density $\rho_b$ in $T_{\mu\nu}^{\rm Universe}$
in equation (\ref{e1}) does not appear in the Newtonian Limit 
nor in the Schwarzschild-de Sitter metric.
The Schwarzschild-de Sitter metric
is constructed with the boundary conditions
that its asymptotic form de Sitter.
For the latter we could also demand the asymptotic form to be the metric
of the cosmological background, i.e. Friedmann-Robertson-Walker
metric \cite{Einstein}. In such a case, $\Lambda$
would eventually enter twice, once in the differential equations
through (\ref{e2}) and the second time through
the boundary condition where $\Lambda$ is part of 
cosmological background metric. This clearly shows
its double role, due to the fact that it is part
of $G_{\mu\nu}$.    

Hence strictly speaking we have to compare the oscillatory part of the solution
with the non-oscillatory one (\ref{solutionLambda}). 
This could be done in a simplistic way taking the
amplitude of the oscillatory part $\hat{h}$ and comparing it with $L^2/r_{\Lambda}^2$ which comes from
(\ref{solutionLambda}). 
This comparison would conceptually not be very different from what we obtained in (\ref{critical}). 
However,
this way the critical distance would come out smaller than (\ref{critical}) (of the order
$\sqrt{\hat{h}}r_{\Lambda}$) as  (\ref{critical}) is suppressed in addition by $fr_{\Lambda}$.
It makes therefore more sense
to make a more sophisticated analysis as done in section 2. If the distance of origin of the
gravitational wave is larger than ${\cal L}_{\rm crit}$, the non-oscillatory background is larger 
than the  
actual wave and therefore the wave interpretation untenable. 

In \cite{Wheeler} a condition for the validity of the picture of a gravitational wave as a ripple
on spacetime is given. Essentially it states that the wave length must be much smaller then
the curvature background ${\cal R}$. Since the part of solution which is connected to $\Lambda$
is non-oscillatory we cannot make such a direct test. In case of a non-zero graviton mass $m$
trigonometric functions $\cos(mt)$, $\sin(mt)$ would enter the solution (\ref{solutionLambda}) 
as shown in \cite{dvali}. Then, the above criteria would apply. In our case, we could compare
the change of $\xi_{\mu \nu}$ by calculating $\delta \xi_{\mu \nu}L/{\cal R} 
\sim {\cal O}(L^3/r_{\Lambda}^3)$ which is much smaller than one as long as $L$ is 
of astrophysical order of magnitude. The reason why the oscillatory solution becomes
comparable to the non-oscillatory is 
because the amplitude of the oscillatory is small.
\subsection{LOCAL VERSUS GLOBAL TEST}
The central point of our interpretation is that given any modified Einstein 
tensor, there has to be {\it two} tests of the gravitational waves:
\begin{itemize}
\item[1.]
{\it Global cosmological test:} 
This test can be presented without any recourse to the 
details of the modified Einstein's tensor. The test consists in {\it
global comparison} between a given background cosmology and the
wave solution $h_{\mu \nu}^{\rm wave}$
of the gravitational wave. It is crucial to realize that we are comparing
here only part of the full solution of the gravitational wave, namely the
wave part $h_{\mu \nu}^{\rm wave}$ (in the case of $\Lambda$ we called it 
$\gamma_{\mu \nu}$). 
This pre-assumes, however, that the wave part is the 
{\it dominant part} of the full local solution 
$h_{\mu \nu}=h_{\mu \nu}^{\rm wave}+h_{\mu \nu}^{\rm rest}$
 where the $h_{\mu \nu}^{\rm rest}$ is due to the modifications
in the Einstein's tensor ($\xi_{\mu \nu}$ in our specific case).
Such a comparison of $h_{\mu \nu}^{\rm wave}$ with the cosmological
background is exactly the Misner-Thorne-Wheeler method \cite{Wheeler} 
mentioned above. One does not compare
$h_{\mu \nu}$ or $h_{\mu \nu}^{\rm rest}$ with the cosmological background, but
only the wave part $h_{\mu \nu}^{\rm wave}$ such that all quantities 
needed for such a test like wavelength, background curvature etc are well 
defined.  The cosmological constant enters here only through cosmology i.e
equation \ref{e1}. In the present paper we are not concerned about this global 
cosmological test. We rather assume that all sources for LIGO and LISA passed 
this test already.

There will be also other effects 
whose root can be traced back to the cosmological background.
One of them will be the direct appearance of such a background in the
propagation of gravitational waves for large distances. This effect
can be taken into account by expanding the Einstein's equations
around the de Sitter metric \cite{dS1}. Such a procedure to include cosmological effects
is not general (indeed a general procedure does not exist), 
but is for the present epoch of the universe which is dominated
by $\Lambda$. The most important effect is the exponentially decay of the wave \cite{dS2}.
As long as $r \ll r_{\Lambda}$ 
(or $ T \ll T_{\Lambda}= r_{\Lambda}$) we can, however, still rely on the 
expansion around the Minkowski metric.
\item[2.] 
{\it Local test:}
In the {\it global test} above we made the assumption that 
$h_{\mu \nu}^{\rm wave}$ is the dominant part of the full solution.
This has to be done in a more
quantitative manner i.e. we have to device  a {\it local test}
which will decide when the wave character is dominant. In this 
{\it local test} $\Lambda$ enter through equation (17). 
Our suggestion for such a test 
is based on the power $P$ as explained in the text above.
This test is rather conservative as other, more `naive' tests yield a smaller 
${\cal L}_{\rm crit}$. 

For a better understanding the difference of the {\it two} tests, let us
visualize a universe (or, equivalently, an earlier epoch of our universe) where
the cosmology is dominated by the background density and not $\Lambda$.
We could then drop $\Lambda$ in (16), but not in (17). The appearance and
relevance  of
$\Lambda$ in the local test would pertain i.e. 
the results of such a test would be the same in any epoch of the universe or any universe with the same
$\Lambda$ (and different background density).
\end{itemize}
\begin{table}[h]
\caption{\label{tab2} Sources of gravitational waves for LISA from references \cite{lisaobs}. 
The last entry is a 
special white dwarf binary. For $P$ we have used geometrized units $G=c=1$.}
\begin{tabular}{|l|l|l|l|l|l|l|}
\hline
System 
& f [Hz] & $\hat{h}$
&Distance & ${\cal L}_{\rm crit}$& $dP/d\Omega$& 
$dP/d\Omega$\\
 & &  & [pc] & [pc]& $\Lambda=0$& $\Lambda \neq 0$\\ 
\hline
\hline
X-ray pulsar binary & $7.9\times 10^{-4}$ &$6\times 10^{-24}$  
&$8000$ &61  &$-$ &$ -$ \\ 
4U1626-67 & & & & & &\\
\hline
X-ray pulsar binary &$3 \times 10^{-3}$  &$2\times 10^{-23}$  &$8100$ &773  &$-$ &$-$   \\
4U1820-30 & & & & & &
\\ \hline
White dwarf binary &$1.4 \times 10^{-4}$  &$2\times 10^{-22}$  &$100$ &$360$  &$1.3 \times 10^{-31}$ 
& $1.2 \times 10^{-31}$\\
WD 2331+290 & & & & & &  \\ \hline 
White dwarf binary &$1.6 \times 10^{-4}$  &$2\times 10^{-22}$  &$100$ &$412$  &$1.7 \times 10^{-31}$ 
& $1.6 \times 10^{-31}$\\
WD 1101+364 & & & & & &  \\ \hline 
White dwarf-B star &$2.4 \times 10^{-4}$  &$1\times 10^{-21}$  &$100$ &$3090$  &$9.57 \times 10^{-30}$ 
& $9.56 \times 10^{-30}$\\
KPD 1930+2752 & & & & & &  \\ \hline 
RX$\dot{\rm J}$080 &$6.2 \times 10^{-3}$  &$4\times 10^{-22}$  &$300$ &$32\times 10^3$  &$9.2 \times 10^{-27}$ 
& $9.2 \times 10^{-27}$\\
6.3+1527  & & & & & &  \\ \hline 
\end{tabular}
\end{table}   
\section{Conclusions}
Any gravity theories with modified Einstein's tensor will require a re-interpretation
of the picture of gravitational waves as ripples on spacetime. The solution of the
linearized new Einstein's equations will contain the oscillatory part (wave) plus
a new component associated with the extension of Einstein's tensor. The latter will not be
oscillatory, in general. It is clear that a suitable comparison between these
two solutions is due in order to be able to say when the wave picture can be maintained.
In this paper we suggested such a comparison by using the gravitational energy momentum tensor
associated with observables. Applying the method for a theory with the cosmological constant,
we deduced a maximal distance beyond which the wave picture loses its meaning.
This makes a direct connection between gravitational waves and theories with $\Lambda$
explaining the accelerated universe.

\renewcommand{\theequation}{A-\arabic{equation}}
\setcounter{equation}{0}  

\section*{Appendix: The Veltman Lagrangian}
It is instructive to re-derive the same linearized equations as in (\ref{lineareq}) and to 
cast a brief glance at the reason why the term proportional $\Lambda$ is not dynamical i.e.
proportional to $\eta_{\mu \nu}$. For this purpose we evoke the Lagrangian given by Veltman in
\cite{Veltman} which reads
\begin{eqnarray} \label{Lag1}
 {\cal L}_h=-2\Lambda\left(1+\frac{1}{2}h-\frac{1}{4}h_{\alpha \beta}h^{\alpha \beta}
+\frac{1}{8}hh\right)-\frac{1}{4}\partial_\nu h_{\alpha \beta}\partial^\nu h^{\alpha \beta}
+\frac{1}{4}\partial_\mu h\partial^\mu h-\frac{1}{2}\partial_\beta h\partial_\mu h^{\beta \mu}\\\nonumber
 +\frac{1}{2}\partial_\alpha h_{\nu \beta}\partial^\nu h^{\alpha \beta}
\end{eqnarray}
The part proportional $\Lambda$ is not gauge invariant under $h_{\mu \nu} \to h_{\mu \nu}
+\partial_{\mu}\epsilon_{\nu}+\partial_{\nu}\epsilon_{\mu}$. Indeed, one obtains under this transformation
\begin{multline} \label{gauge}
 2\Lambda\left(1+\frac{1}{2}h-\frac{1}{4}h_{\alpha \beta}h^{\alpha \beta}+\frac{1}{8}hh\right)\to\\
 2\Lambda\left(1+\frac{1}{2}h -\frac{1}{4}h_{\alpha \beta}h^{\alpha \beta}+\frac{1}{8}hh
+ \partial^\sigma \epsilon_\sigma-h_{\alpha \beta}\partial^\alpha \epsilon^\beta
 +\frac{1}{2}h\partial^\sigma\epsilon_\sigma\right)
\end{multline}
The formal condition for the gauge invariance to hold up to total derivative is obviously
\begin{equation} \label{condition}
h_{\alpha \beta}\partial^\alpha\epsilon^\beta=\frac{1}{2}h\partial^\sigma\epsilon_\sigma
\end{equation}
The correct gauge invariant Lagrangian is simply 
\begin{equation} \label{correct}
{\cal L}_h^{\prime}={\cal L}_h +2\Lambda (1/8 hh -1/4 h_{\alpha \beta}
h^{\alpha \beta})
\end{equation}
In vacuum, the Euler-Lagrange equations according ${\cal L}_h$ come out to be
\begin{equation}   \label{EL}
\square h^{\mu \nu}-\eta^{\mu \nu}\square h+\eta^{\mu \nu}\partial_{\sigma}\partial_{\alpha}h^{\sigma \alpha}
+\partial^\mu \partial^\nu h-\partial_\sigma \partial^\mu h^{\sigma \nu}
-\partial_\sigma \partial^\nu h^{\sigma \mu}=2\Lambda \eta^{\mu \nu}
-2\Lambda \left(h^{\mu \nu}-\frac{1}{2}\eta^{\mu \nu}h\right)
\end{equation}
The second term on the right hand side of (\ref{EL}) is due to the the non-gauge invariant
terms in the Lagrangian. Dropping this term results 
in equation of motion which we had before i.e. (\ref{lineareq}). 
(this is equivalent to use the gauge invariant Lagrangian (\ref{correct}). 
Indeed, taking the trace of  
\begin{equation}   \label{EL2}
\square h^{\mu \nu}-\eta^{\mu \nu}\square h+\eta^{\mu \nu}\partial_{\sigma}\partial_{\alpha}h^{\sigma \alpha}
+\partial^\mu \partial^\nu h-\partial_\sigma \partial^\mu h^{\sigma \nu}
-\partial_\sigma \partial^\nu h^{\sigma \mu}=2\Lambda \eta^{\mu \nu}
\end{equation}
and multiplying with $\eta_{\mu \nu}$ we can replace 
$-\eta_{\mu \nu} \square h= -\eta_{\mu \nu}\partial_{\sigma}\partial_{\alpha}h^{\sigma \alpha}
+ 4 \Lambda \eta_{\mu \nu}$ in (\ref{EL2}) to arrive at (\ref{lineareq}) in vacuum ( with matter
the steps to obtain (\ref{lineareq}) are similar).
This shows once again that equation (\ref{lineareq}) is correct.

In passing let us make a comment regarding the gauge invariance of (\ref{Lag1}).
Taking the divergence of equation (\ref{EL}) gives us
\begin{equation}   \label{eq:De Donder}
\partial^\mu h_{\mu \nu}=\frac{1}{2}\partial_\nu h
\end{equation}
which is actually the de Donder condition, now not as a gauge fixing, but as a
result of the equation of motion (this is in analogy to the free
massive vector case $A_{\mu}$ where in spite of the loss of gauge invariance
the equation of motion gives us the Lorentz gauge $\partial_{\mu}A^{\mu}=0$).
If we put this equation into the Lagrangian (\ref{Lag1}), then according to
(\ref{gauge}) and (\ref{condition}) the total Lagrangian would come out now
gauge invariant up to total derivatives. Obviously, this is in contradiction with our previous
result and the resolution of this seemingly different results is that it is
not permissible to use equations of motion (or a part of them) in the
Lagrangian itself. Similarly, we cannot use a gauge fixing in (\ref{correct})
without changing physical results. For instance, if we use the traceless
gauge $h=0$ in (\ref{correct}), the term $(1/4)\partial_{\gamma}h
\partial^{\gamma}h$ in (\ref{correct}) would be absent.
Such a term under variation of the action gives $(1/2)\eta_{\mu \nu}\square h$
which turns out to be crucial to obtain the equation (\ref{lineareq}) as explained above.


\begin{thebibliography}{99}
\bibitem{footnote}
As usual we have $g^{\mu \nu}=\eta^{\mu \nu} -h^{\mu \nu}$ such that
$g^{\mu \alpha} g_{\mu \beta}= \delta^{\alpha}_{\beta} + {\cal O}(h^2)$ which is sufficient
for the order of expansion at which we are working (first order in $h_{\mu \nu}$). Had we
further expand $h_{mu \nu}$ according to inverse powers of the velocity of light
(see \cite{Brumberg})  we would have to pay more attention to this relation (see
\cite{weinberg}). This also shows that the mixed tensor form of Einstein tensor
i. e. $R^{\alpha}_{\beta}-(1/2)R\delta^{\alpha}_{\beta}+\Lambda \delta^{\alpha}_{\beta}$,
where the delta-function comes from the {\it full} contraction of the metric tensors, is not really
a suitable form for the perturbation of higher orders.
\end{thebibliography}
\end{document}